# From a set of parts to an indivisible whole.
# Part III: Holistic space of multi-object relations

Leonid Andreev

Equicom, Inc., 10273 E Emily Dr, Tucson, AZ 85730, U.S.A.
E-mail: equicom@matrixreasoning.com

## Abstract

The previously described methodology for hierarchical grouping of objects through iterative averaging has been used for simulation of cooperative interactions between objects of a system with the purpose of investigation of the conformational organization of the system. Interactions between objects were analyzed within the space of an isotropic field of one of the objects (drifter). Such an isotropic field of an individual object can be viewed as a prototype of computer ego. A drifter's isotropic field allows visualization of a holistic space of multi-object relations (HSMOR) which has a complex structure depending on the number of objects, their mutual arrangement in space, and the type of metric used for assessment of the objects' (dis)similarities. In the course of computer simulation of cooperative interactions between the objects, only those points of the space were registered which corresponded to transitions in hierarchical grouping. Such points appeared to aggregate into complex spatial structures determining a unique internal organization of a respective HSMOR. We describe some of the peculiarities of such structures, referred to by us as attractor membranes, and discuss their properties. We also demonstrate the peculiarities of the changing of intergroup similarities upon a drifter's infinite distancing away from the fixed objects.

**Keywords:** Iterative averaging, computer ego, multi-object relations, cooperative interactions, system's conformational organization, attractor membranes, aura, holism, metrics

## 1. Introduction

The phenomenon of iterative averaging (IA) previously described by us in [1, 2] provides a capability to radically transform any of the currently known information-processing technologies. The IA phenomenon consists in the fact that any system subjected to iterative averaging undergoes a division that produces two alternative subgroups of elements, without outliers. There are two key points about the IA phenomenon. Firstly, any (dis)similarity matrix subjected to at least one act of averaging turns into a closed-type system. Upon further transformations, the system, being completely isolated from all the diversity of the outer information evolves only in the direction of establishment and development of the intra-system relations in accordance with a mechanism directed towards the abstraction of those relations. Secondly, due to spontaneous, interconnected and interdependent processes occurring in the course of iterative averaging, the objects of a system under processing interact with each other based on the principles that are similar to the principle of self-organization in nature and society. These two peculiarities of the IA phenomenon create a capability to discover natural hierarchical structures of systems under analysis, which are determined solely by input datasets and by no means depend on a data analyst's decisions. This property of the IA phenomenon is very important upon the use of the IA method as a platform for various methodological approaches to a wide range of problems related to the concept of artificial intelligence. As was pointed up by V. Estivill-Castro [3], "... there are many clustering algorithms, because the notion of "cluster" cannot be precisely defined. ... Researchers have proposed many induction principles and models whose corresponding optimization problem can only be approximately solved by an even larger number of algorithms". Therefore, the IA method is fundamentally different from any of the clustering methods as it ensures, as early as at the very first step of a system processing, that the relations between all of the elements of the system non-alternatively turn into interdependency based on objective principles of a physical nature. The induction principles, on which the classical science is based, have absolutely no relevance for the character of such relationships.

The IA algorithm is the only adequate tool for decomposition of a complex system into a set of subgroups joined by strictly hierarchical relationships. It can also be used as an autonomous engine that can be incorporated into any kind of data processing procedure as an element performing, independently and in unsupervised mode, a certain autonomous function that is not dependent on other operations involved in the overall procedure. An example of IA-based autonomous computing is the infothyristor previously described as part of the HGV2C method [4] providing a universal technology for pattern recognition wherein the infothyristor autonomously performs a subdivision of a set of three components into two subgroups based on the IA principle. The said technology for pattern recognition provides a capability to evaluate any particular pattern of data point distribution from the "standpoint" of one of the objects of the system or an exogenous object constructed based on a certain hypothesis (model).  As was previously demonstrated [2], evolution of a system's complexity under the effect of iterative averaging has qualitatively similar dynamics, whereas quantitatively it may significantly vary depending on the number of elements in a system and the values of parameters describing the elements: the achieving of a stable final result may require just a few IA cycles for some systems, or hundreds for others. However, the overall similarity of the dynamics of IA of different systems by no means indicates the predictability of a final result for a particular system: the way of how a given totality of objects will split into subgroups cannot be predicted through either theoretical calculations or practical assessment. Another factor that further contributes to the unpredictability of a final result of the IA processing is the high cooperativity inherent in the iterative averaging process: an addition of even one new element to a system of hundreds elements, or a removal of at least one of the existing elements can result in drastic changes in the composition of subgroups emerging upon the IA process. Fortunately, these two factors that are not favorable for the matter of scientific formulation and practical realization of the IA approach are more than compensated by a remarkable advantage provided by the IA processing: all of the IA-based analyses conducted by us on numerous data systems invariably demonstrated a highly intelligent potential of the IA procedure. Some of the examples were provided in our previous publications: analysis of scattered points, analysis of climatic data involving 108 parameters for 100 cities, demographic data analysis [2], human pose recognition [5], etc. It would be more important to state that in the course of the many years of working with the IA method, we have never encountered any senseless results. The non-biological intelligence clearly expressed in the IA phenomenon certainly needs further multi-lateral and objective investigations involving as many as possible practical examples of analysis, which can be done with the use of *MeaningFinder* (Equicom, Inc.), a software implementation of the IA-based information processing technology. However, there are also other, fundamentally different and more in-depth, approaches to investigation of the IA phenomenon. One of them is discussed in this paper.

## 2. Computer ego

Throughout the history of human civilization, few areas of high academic and applied interest have attracted as much funding as the attempts of development of artificial intelligence. The fact that all of the efforts in this area of computer science which have been undertaken for more half a century have so far delivered almost nothing that could really be called 'intelligence' is not that striking – after all, this is a task of tremendous complexity and it requires long-term collective efforts. However, what strikes about the ongoing research and development efforts in this area is the fact that they all ignore a well-known truth that there is no intelligence without ego. The conceptual problem of computer ego has never been tackled as a potential solution to the problem of artificial intelligence. 'Ego' has many definitions, and the philosophical understanding of its relation with mind, intelligence, self, consciousness, thought, intuition, knowledge, etc. has a long and complex history involving many great thinkers, such as Kant, Heidegger, Schelling, Fichte, Freud and others. Our task within the context of the described technology was much simpler and more utilitarian: how can computer intelligence benefit from having a computer ego?

In humans, loss of ego leads to depersonalization of intellect, i.e. loss of normal sense of reality. Computers have never had the ego and, therefore, cannot lose it, but an interesting question is the reverse: what would happen if a computer could acquire the ego? As far as the form of functioning of the computer ego (CE) is concerned, there are several possible variants. (1) Unlike humans who, in the norm, have one ego that is unique for a given individual, the CE can have an unlimited number of hypostases, and the computer can be switched to ego-1, ego-2, etc. modes. (2) Unlike the human ego that, in the norm, is wholesome, the CE can have a mosaic structure, combining seemingly incompatible elements. (3) Unlike the human ego, the CE can be totally positive or totally negative in relation to a certain object or event, and the two alternative egos can be engaged as necessary. Apparently, that form of CE could be used in World Wide Web technology, web design, construction of search engines, etc. There may be many other forms of functioning of the CE. As to methodology of CE application, the CE could be used both for element-by-element reductionism analysis of systems, and for supra-system and intersystem investigations from the standpoint of holism. The IA method can serve as a standard and universal analytical method in such investigations. The development and testing of the principles of the CE utilization in computer intelligence has been our strategic task.

## 3. Use of computer ego in analysis of system organization

The IA effect on sets of scattered points [1-2, 4-5] convincingly demonstrates that the IA processing leads to the emergence of a certain organization of the system under processing, which fully conforms to the system criteria according to the theory of general systems [6] which views a system as a sets of elements together with the set of the relations among them [7]. As was emphasized by V. Majernik [8], "a general system is not an aggregation of some objects. Rather it is set of interrelated, intercooperating or interconflicting parts creating through their interaction new system properties which do not exist in a loose collection of individual objects". System organization processes that spontaneously occur in nature and society are investigated by both theoretical and applied sciences not only and not so much in the direction of understanding of the emerging configurational and functional organizations, but mostly in an attempt of solving a more pragmatic problem: determination of which of the variables of a system's elements and which of the elements are most responsible for the emergence and occurrence of the system organization. This approach is fundamentally wrong as the contribution of either the elements or their variables to a real process of system-organization cannot be established through reductionism. Isolating an individual element of a system in order to analyze that element's contribution to the system is senseless as the system itself, without that element, is no longer the same system as it was before the removal of one individual element. Despite the great variety of the methods of study of system organization, they can provide only indirect and incidental knowledge, as this problem cannot be solved outside of the paradigm of holism. Watanabe [9] proposed an entropy criterion for assessment of systems' configurational organization, which represents the sum of entropies of the parts of system minus entropy of the whole system. Although logical and formally valid, this approach clearly cannot be used in practice. Even a detailed picture of a system's hierarchical structure which can be obtained by the use of the holism-based IA method is not sufficient for understanding of the configurational structure of interrelations between the individual elements of a given system. Whatever kinds of relationships may be responsible for intercooperational and interconflicting processes in system organization, they always have a finite spatial range of operation and are non-uniform in their operation within a space determined by various variables. In this paper, we describe a new approach to analysis of configurational organization of systems which is based on the IA method and provides quite a simple technique for visualization of a holistic space of multi-object relations (HSMOR).

The said technique is based on the idea of computer ego (CE) [4, 5] and can be used for analysis of conformation structure of system organization. Previously [4], we described how CE can be used as a test object for assessment of separate, non-interacting individual objects. The methodological solution provided in this paper allows the assessment of interactions between objects within a system and is based on the idea of using the CE that is represented not by an individual object but by the isotropic field of an individual test object. The intra-system relations of the test object with the rest of the system's objects are assessed at each point of its isotropic field. For the purposes of practical realization of this idea, the isotropic field of an object can be visualized as a space in which a chaotically and unrestrictedly moving test object, hereunder referred to as 'drifter' ("Dr"), comes into cooperative interaction with other objects whose location is fixed (referred to as fixed objects, "FO"). Interactions between Dr and FO simulated by the IA algorithm eventually result in a system organization.

## 4. Holistic space of multi-object relations (HSMOR) between a drifter and two fixed points

Let us consider a simplest variant of system organization involving three elements: a drifter and two fixed objects. For the purpose of easier visualization of the system's configurational organization that will emerge as a result of the IA processing, we will use only three parameters describing the Dr and FO, even though the IA-method does not restrict the number of parameters.

The term 'cooperative interactions' is widely used in research publications (see, e.g. [10-11]), and its general meaning is certainly clear; however, its exact attribution has neither been defined in the modern science, nor have there been any attempts of defining it, and therefore it has been liberally used in the meaning that directly follows from the word combination. The term 'cooperative interactions' as is used in contemporary science helps to avoid some of the difficulties caused by certain gaps in the modern fundamental scientific knowledge, and it conveniently masks the lack of understanding of the fine mechanisms that are involved in system organization. The general properties of those mechanisms are the same in different systems, even though the interactions between the elements may be determined by attraction and repulsion forces of different physical nature. The results presented in this paper, as well as those yet to be reported in our following publications, provide a better understanding of the phenomenon of 'cooperative interactions' which, as it appears, can even be quantitatively assessed.

In this work, we used a number of different metrics for computation of (dis)similarity matrices for Dr and FOs. The advantage of such an approach is based on the fact that different metrics differently reflect the nature of interactions between the elements of a system under formation, and, therefore, both similarities and dissimilarities between the discovered regularities may appear to be a source of useful information. We used three metrics: (1) Euclidean distances (ED), (2) city-block (CB), and (3) XR-metric (B = 1.50). The first two metrics are commonly used in computation of dissimilarity coefficients [12]. The XR-metric proposed by us in [1-2, 5] represents a criterion that summarily reflects a difference in a parameter values through the use of an exponential function, and, therefore, is referred to as a shape-metric. The ED metric was applied by us in a conventional way: for calculation of distances between the points based on the x, y and z values according to Pythagorean Theorem. Matrices of dissimilarity (based on CB) and similarity (based on XR) were computed by the method of similarity matrix hybridization described in [2, 13], which involves computation of monomer matrices according to each individual parameter, and the following hybridization of the monomer matrices by computing geometric means for each set of identical cells of the monomer matrices. By applying the IA-method to each of the (dis)similarity matrices, the set of points (a drifter and fixed points) was subdivided into a maximal number of subgroups. On the resulting diagrams, we registered only those points that were on the borderline between two kinds of grouping, i.e. points of transition.

Fig. 1 shows the results obtained with the use of Euclidean distances as a metric, for Dr and two FOs, A and B, whose coordinates $x$, $y$, and $z$ had values of 1, 1, 0 and 0, 0, 1, respectively. There are three possible variants of subdivision: 'AB – Dr'; 'A – BDr'; and 'B – ADr'. Fig. 1 shows all three variants and demonstrates a number of phenomenal peculiarities: (1) Despite the fact that the coordinates of Dr vary within the range of −∞ to + ∞, the interactions between Dr and the two FOs are restricted in space and occur within an area whose shape is considerably more complex than the spatial structure made of points A and B. (2) Outside of the space shown in Fig.1, there can only be one variant of subgrouping: 'AB – Dr'. (3) The space of interactions of Dr with two FOs, whose size is 1, exceeds the latter by 6.4 times, i.e. fits a cube whose edge has a size of 6.4. (4) The width of the transition borderline between the variant 'AB – Dr' and two other variants is infinitesimally small. (5) Even with a simplest set of points – such as two fixed points – the configuration of the system of 'Dr – 2xFO' cannot be visualized through a 3D HSMOR as various parts of the areas of subgrouping 'A – BDr' and 'B – ADr' are overlapping. We will discuss this issue further in this paper.

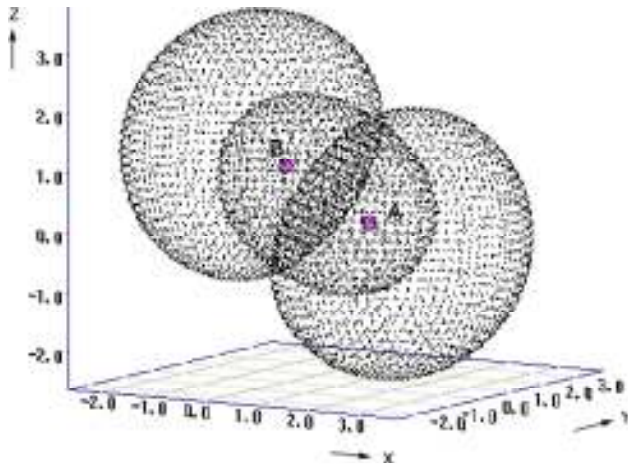

Fig.1. Holistic space of multi-object relations, HSMOR(ED), between a drifter and two fixed points, A and B, computed by using Euclidean distances. The values of coordinates $x$, $y$, and $z$ of points A and B shown as large red dots are: 1, 1, 0; and 0, 0, 1, respectively.

Fig. 1 shows five constituent parts of the resulting HSMOR figure: two larger spheres, two smaller spheres, and a lens-shape area that is adjacent to each of the four spheres. The HSMOR figures obtained with the use of metrics CB and XR have significantly more complex configurations. In case of the CB metric (Fig. 2), the HSMOR is represented by two very complex, partially integrated into each other spatial figures with flat surfaces. The HSMOR based on XR-metric has a shape of two partially integrated into each other octahedrons, wherein each octahedron represents a pair of equilateral pyramids.

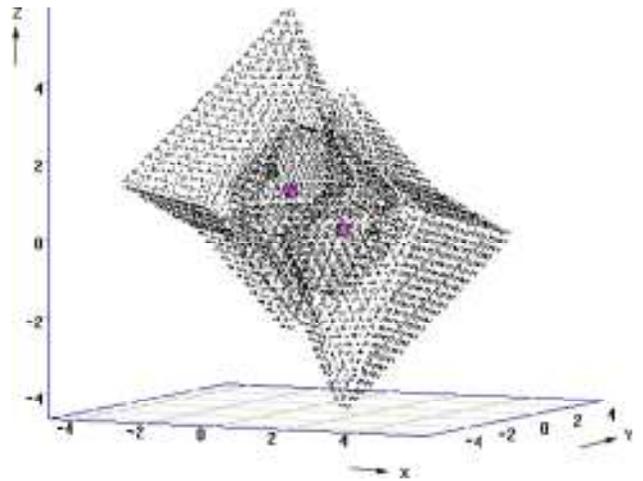

Fig.2. Holistic space of multi-object relations, HSMOR (CB), between a drifter and two fixed points, A and B, computed by using the city-block metric. The values of coordinates $x$, $y$, and $z$ of points A and B shown as large red dots are: 1, 1, 0; and 0, 0, 1, respectively.

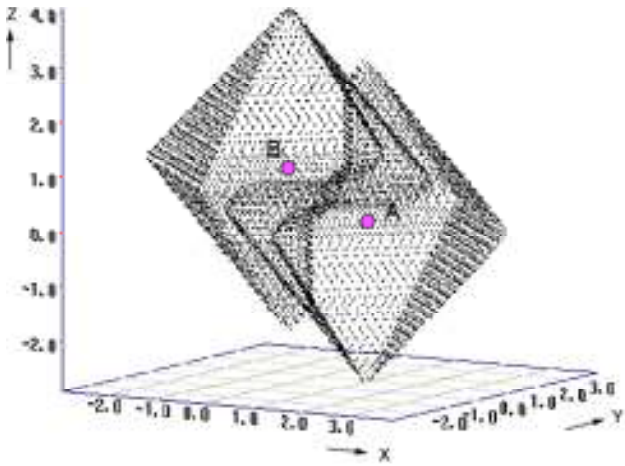

Fig. 3. Holistic space of multi-object relations, HSMOR (XR), between a drifter and two fixed points, A and B, computed by using the XR-metric. The values of coordinates $x$, $y$, and $z$ of points A and B shown as large red dots are: 1, 1, 0; and 0, 0, 1, respectively.

Even though the three variants – HSMOR (ED), HSMOR(CB), and HSMOR(XR) – were computed by using different metrics and drastically differ in configuration, their sizes are within a same order of magnitude: 6.4 (ED), 10.3 (CB), 6.8 (XR).

In order to better understand the internal structure of the HSMOR figures, we made a cross-section of each of them across the $z$-axis. Fig. 4 shows the results of cross-sections at $z = -2$, $-1$, $0.5$, $2$, and $3$, respectively. As is seen, they represent different geometrical figures of a regular form, which are symmetrical relative to the midpoint of the distance between points A and B ($z = 0.5$). As was pointed out earlier, the subgrouping of Dr, A, and B outside of the HSMOR figure (which applies to HSMOR(CB) and HSMOR(XR) as well) is always the same, i.e. Dr – ∑ FO, which should indicate that outside of

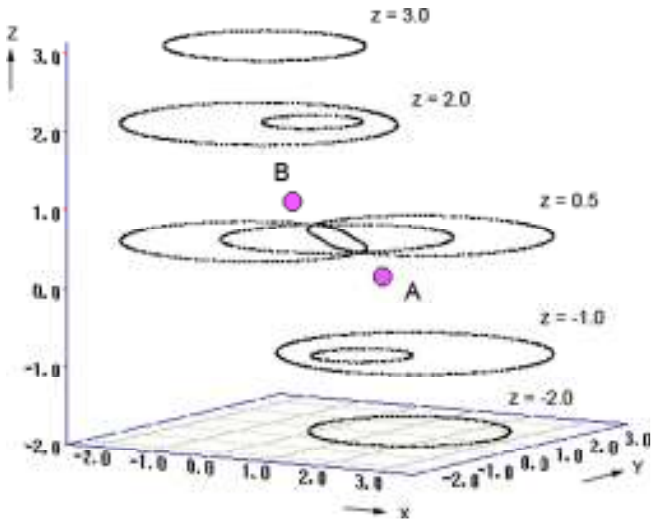
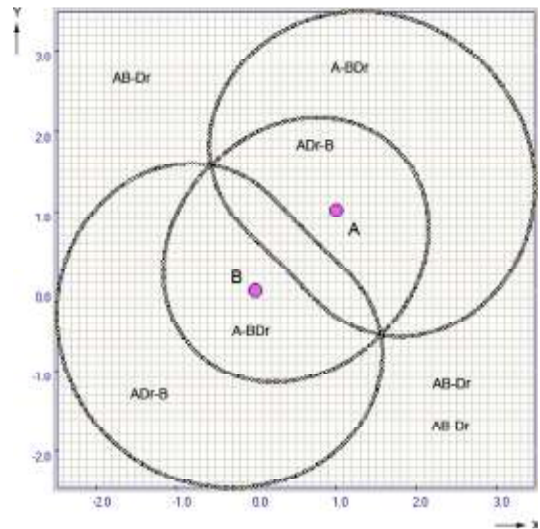

Figs. 4a and 4b. Cross-sections of the HSMOR (ED) across the $z$-axis. 4a (left), cross-sections at $z = -2$, 1, 0.5, 2, and 3, respectively. 4b (right) shows the variants of subgrouping of Dr, A, and B at the cross-section at $z = 0.5$. (The values of coordinates $x$, $y$, and $z$ of points A and B shown as large red dots are: 1, 1, 0; and 0, 0, 1, respectively.)

the HSMOR the drifter stops interacting with the fixed points A and B. Fig. 4b shows the grouping at different areas of the cross-section at $z = 0.5$. Groups 'A – DrB' and 'B – DrA' cover two fragments of the cross-section, each of which consists of larger and smaller circles. In each case, the larger circle corresponds to the closer FO, while the smaller one, respectively, to the farther FO. The lens-shape area between the larger and smaller circles may be a result of mechanical overlapping of the two grouping variants, i.e. 'A – DrB' + 'B – DrA'. The grouping within that area has been found to be Dr – AB.

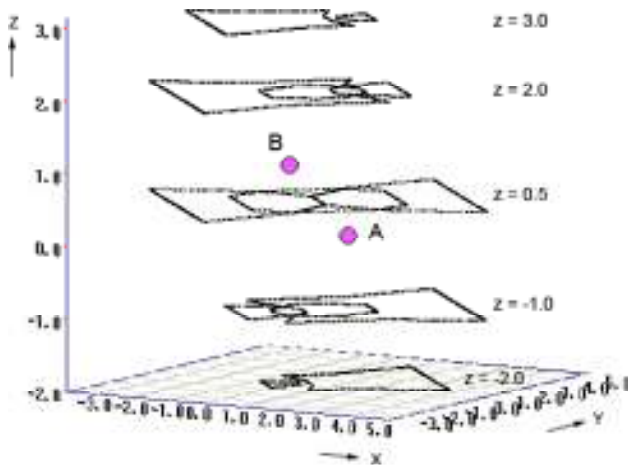
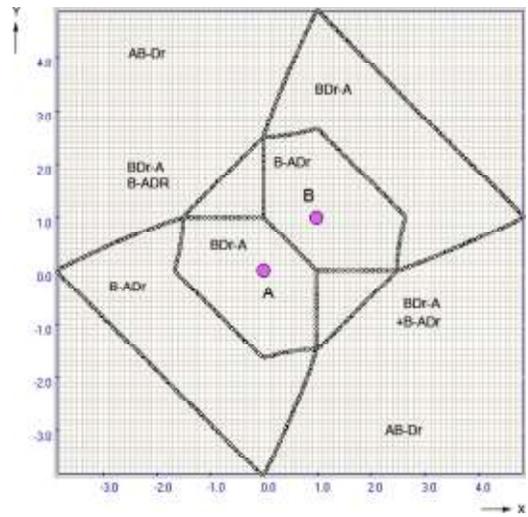

Figs. 5a and 5b. Cross-sections of the HSMOR (CB) across the *z*-axis. 5a (left), cross-sections at z = -2, 1, 0.5, 2, and 3, respectively. 5b (right) shows the variants of subgrouping of Dr, A, and B at the cross-section at *z* = 0.5. (The values of coordinates *x*, *y*, and *z* of points A and B shown as large red dots are: 1, 1, 0; and 0, 0, 1, respectively.)

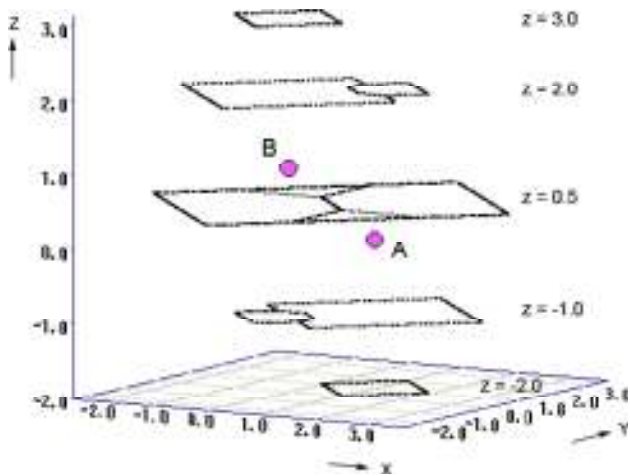
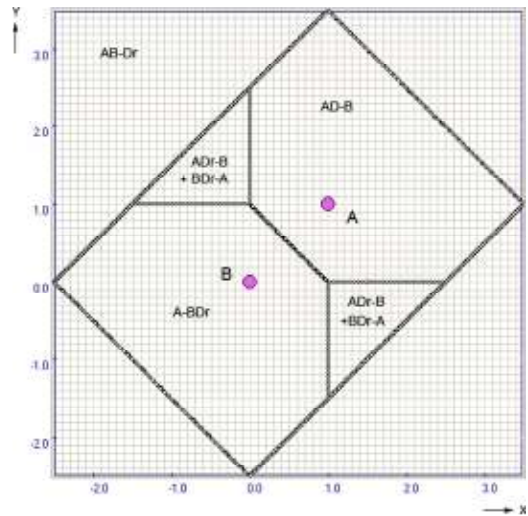

Figs. 6a and 6b. Cross-sections of the HSMOR (XR) across the *z*-axis. 6a (left), cross-sections at *z* = -2, 1, 0.5, 2, and 3, respectively. 6b (right) shows the variants of subgrouping of Dr, A, and B at the cross-section at z = 0.5. (The values of coordinates *x*, *y*, and *z* of points A and B shown as large red dots are: 1, 1, 0; and 0, 0, 1, respectively.)

The above-presented example of interactions between a freely drifting point and two fixed points, A and B, is a truly convincing illustration of the principles of holism – an illustration of the fact that a whole does not equal than the sum of its parts. This is especially well seen on the example of HSMOR(CB). The above-illustrated extremely complex pictures of the interactions between three objects cannot be obtained by any kind of logical manipulations. They are an *a priori* unpredictable result of a very simple operation – the averaging of (dis)similarity coefficients – whose analogs widely occur in both living and non-living nature.

## 5. Holistic space of multi-object relations (HSMOR) for a drifter and more than two fixed points

Upon the increase of the number of fixed points, the number of variants of subgroups emerging in the course of hierarchical grouping, hence the complexity of the internal structure of HSMOR(ED), drastically increases. This is clearly seen on Figs. 7a and 7b that show the HSMOR cross-section at $z = 0.5$ for the set of three fixed points, A, B, C, and D, and a drifter. In this example, points A and B have the same coordinates as in the previous example. All three coordinates of point D have a value of 1.

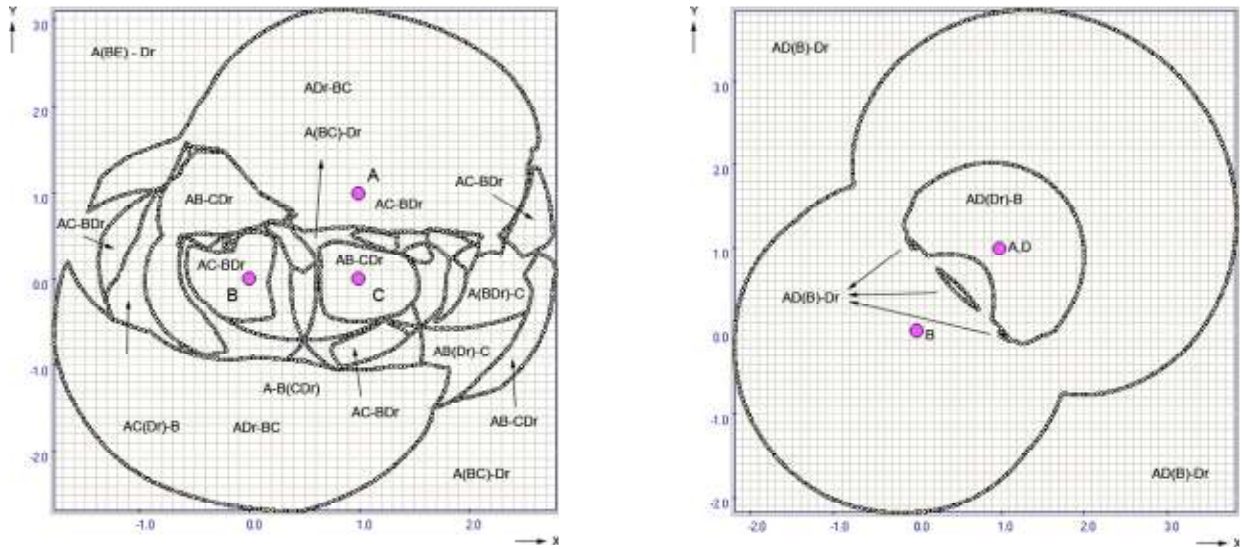

Figs. 7a and 7b. Cross-sections, at $z = 0.5$, of the HSMOR(ED) of sets of three points, A, B, and C (7a, left)) and A, B, and D (7b, right)). Out of three points, A, B, and C, shown as large red dots, points A and B have the same coordinates as in the above examples. The values of coordinates x, y, and z of points C and D shown as large red dots are: 1, 0, 1; and 1, 1, 1, respectively. The figures demonstrate the character of grouping in some of the closed areas of the HSMOR(ED). Hyphens between the symbols indicate subdivisions on a root level, whereas subdivisions on second node levels are shown in parentheses.

Point C differs from point D by the value of $y = 0$. The cross-section of the HSMOR(ED) of points Dr and A, B, and C, shown in Fig. 7a represents an extremely complex picture involving a multitude of closed areas. Each of the closed areas represents a grouping that is different from those in the adjacent areas. In the following examples with more than two FOs, there will be no need in an exhaustive analysis of the character of grouping within each of the areas, as a demonstration of the increase of diversity of grouping caused by the increase of the number of FO's will be a sufficient evidence of the increasing complexity of a respective HSMOR. It should be noted, however, that the configuration of FO's significantly contributes to the complexity of the internal structure of HSMOR(ED). For instance, as is seen from Fig. 7b, in case of the set consisting of Dr, A, B, and D, a "single mutation" – such as a change of the $y$ value of point C from 0 to 1 – results in a drastic decrease of the system's complexity. Complexity of the internal structure of HSMOR figures is determined not only by FO configuration, but also, as was shown earlier, by the number of FO's, as well as by other factors that are hard to predict, at least at the present stage of this study.

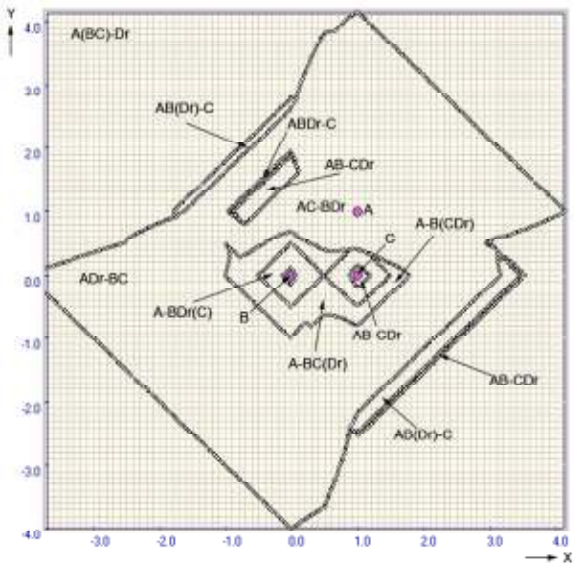 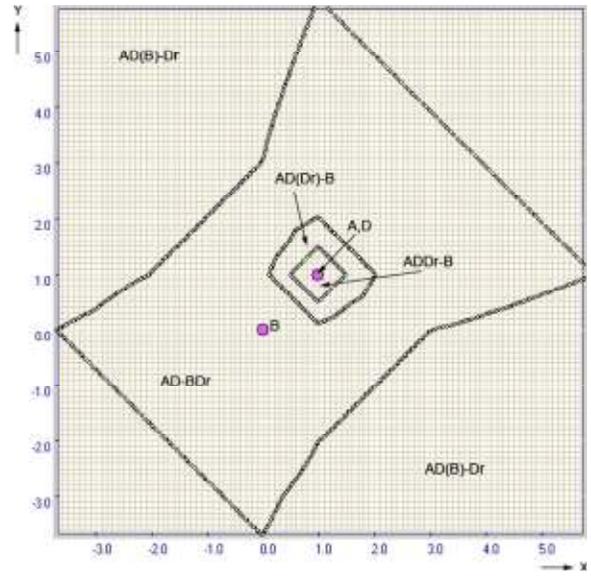

Figs. 8a and 8b. Cross-sections, at $z = 0.5$, of the HSMOR(CB) of sets of three points, A, B, and C (8a, left) and A, B, and D (8b, right). The coordinates of points A, B, C, and D are the same as in Figs. 7. The figures demonstrate the character of grouping in some of the closed areas of the HSMOR(CB). Hyphens between the symbols indicate subdivisions on a root level, whereas subdivisions on second node levels are shown in parentheses.

Of the three metrics used in this experiment, Euclidian distances provide the most complex grouping. A less complex grouping is observed upon the use of CB-metric (Figs. 8a and 8b), whereas the most simple grouping was produced by the XR-metric (Figs. 9a and 9b). A cross-section of the HSMOR(XR) shows mostly linear shape figures.

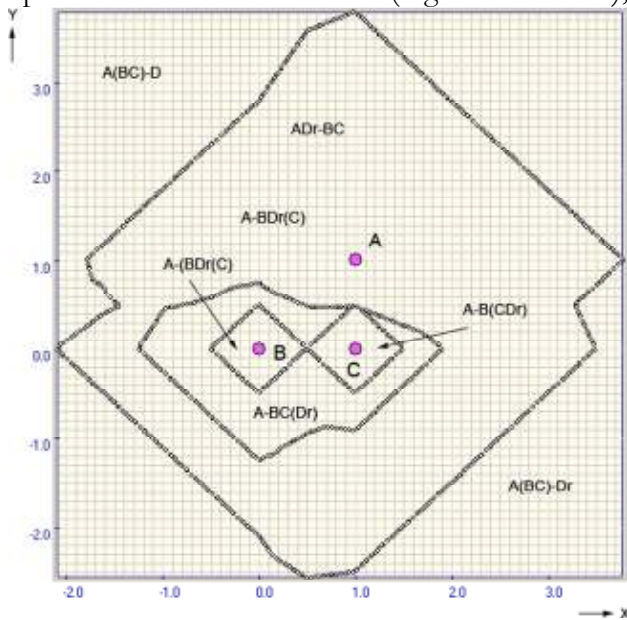 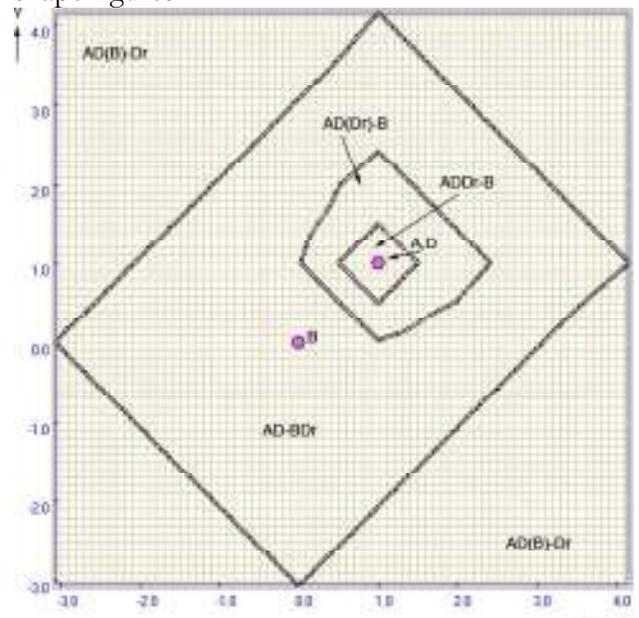

Figs. 9a and 9b. Cross-sections, at $z = 0.5$, of the HSMOR(XR) of sets of three points, A, B, and C (9a, left) and A, B, and D (9b, right). The coordinates of points A, B, C, and D are the same as in Figs. 7. The figures demonstrate the character of grouping in some of the closed areas of the HSMOR(XR). Hyphens between the symbols indicate subdivisions on a root level, whereas subdivisions on second node levels are shown in parentheses.

The role of metrics is an important issue that needs a thorough investigation which should also help to understand the regularities in the changing of similarities between the objects of a complex system upon the changes of distances between the objects. However, in this paper that primarily deals with the fact of the emergence of a spatially structured system of interrelations between a system's objects in the course of iterative averaging of their properties, we will touch upon only the practical aspects of the problem of metrics. Having a basic technology for analysis of interrelations in systems with varying configuration of fixed objects is extremely important. It appears that only the XR-metric allows analysis of interactions of Dr with more than two FOs. Two other metrics – ED and CB – appeared to be totally unsuitable for that purpose. Fig. 10 shows cross-sections across the z-axis of the HSMOR(XR) emerging upon the interactions between Dr and 8 FOs that are positioned as cube corners and have the coordinates 0 and 1. As is seen, the cross-sections made at different $z$ values reveal very complex symmetrical pictures that are drastically different from each other. Figs. 11a and 11b show cross-sections of HSMOR(XR) and HSMOR(CB) at $z$ = 0.5. As is seen, in case of HSMOR(XR), the cross-section of interrelations between Dr and the 8 FOs shows a regular-shape symmetrical picture, whereas the cross-section of HSMOR(CB) has a regular-shape area only in the center where the distances from the drifter to the fixed objects are approximately equal. The rest of the picture is almost chaotic. Upon the use of Euclidean distances, cross-sections of HSMOR show an entirely chaotic picture. Apparently, the regularities in the changing of similarities have the exponential character, and, therefore, the XR-metric is the only suitable metric for analysis of large numbers of fixed objects. There is one more important detail about the above-provided illustrations. A cross-section of HSMOR(XR) for 8 FOs at z = 0.5 is shown in both Fig. 10 and Fig. 11a; however, in case of Fig. 11, the density of scanning of the interactions between Dr and FOs was twice higher than it was in case of Fig. 10. As is seen, the cross-section shown in Fig. 11 provides considerably more details, however, without changing the overall picture of the cross-section shown in Fig. 10. In case of the use of ED and CB metrics, the increase of the scanning density does not result in a more detailed picture of the cross-section but makes it even more chaotic. This provides one more indirect evidence to the effect that the changes in similarities between objects within a space occur exponentially.

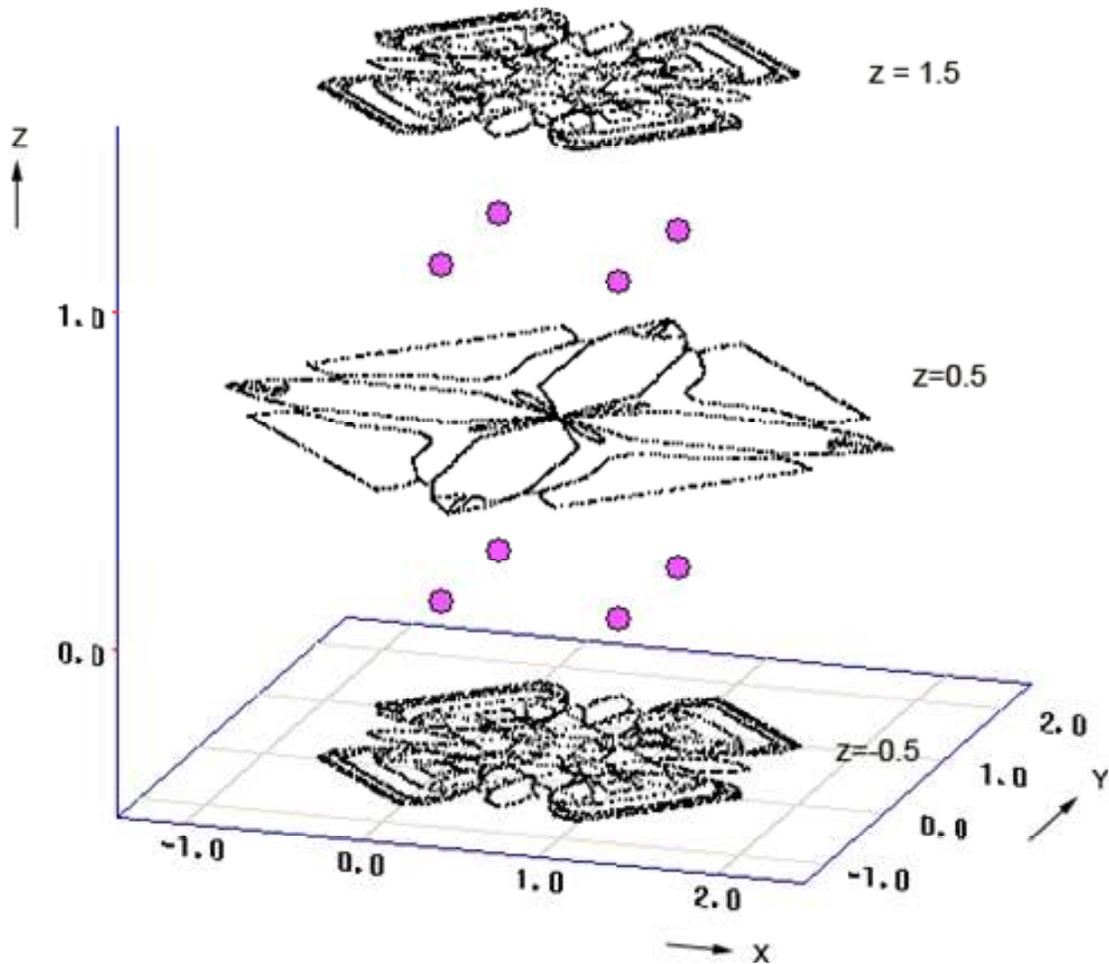

Fig. 10. Cross-sections, at $z$ = 1.5, 0.5, and -0.5 of the HSMOR(XR) emerging upon the interactions between Dr and 8 FOs that are positioned as cube corners with the coordinates 0 and 1.

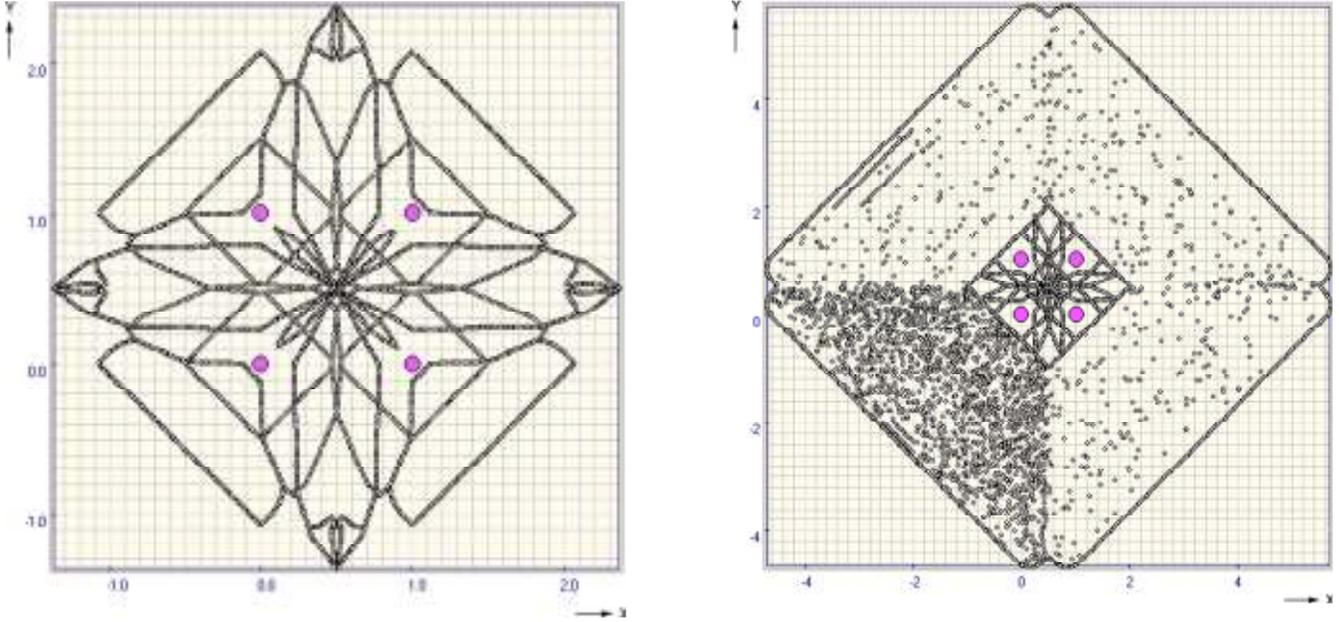

Figs. 11. Cross-sections across the z-axis, at z = 0.5, of HSMOR emerging upon the interactions between Dr and 8 FOs that are positioned as cube corners with the coordinates 0 and 1. Fig. 11a (left), HSMOR(XR); Fig. 11b (right), HSMOR(CB).

## 7. Attractor membrane systems

We have presented a few examples of interactions between a drifter's isotropic field and a set of fixed objects. The use of the iterative averaging algorithm [2] provided simulation of the formation of systems whose conformational organization is unique for any initial set of fixed objects. For visualization of end results, we used only three dimensions, but there is absolutely no problem in dealing with a high-dimensional space of system formation, including a drifter's isotropic field within that space. However, the spatial configuration of HSMOR is typically so complex that even for a 3D visualization of the results obtained with only a few FOs, one needs to either make cross-section at individual coordinates or view only limited areas of HSMOR.

It seems that the isotropic field of a drifter, along with a resulting hierarchical grouping, could be considered as a simplest imitation of one of the important aspects of the mechanism of functioning of consciousness. When we are trying to make a multiple-choice decision and are evaluating the possible choices, we compare the future consequences of each of the choices, i.e. compare multitudes of parameters (consequences) of the choices. We engage in a similar process when we are deciding on making a purchase, or planning a vacation, etc. However primitive the above-discussed model of an object's isotropic field may be, the new discoveries resulting from the use of that model are too important and, once verified on simple examples, should be tried to be extrapolated to more complex objects. For instance, no physical laws and no logical construct can lead to realization of a complex domain character of relations between a drifter's isotropic field and the fixed objects, as it is clear from even the most simple example of a two-FO system demonstrated above (see Figs. 1 – 3). This kind of fundamentally new regularities can be seen on any of the above-provided illustrations.

By definition, in the drifter's isotropic field, we register only those points that correspond to transition between different kinds of grouping of the drifter and fixed objects. However, it is clearly seen that these borderline points always aggregate into wholesome closed systems whose configuration depends, in a very intricate and indirect way, on the values of parameters describing the fixed objects, as well as on the number of the fixed objects. Even in the simplest example with two FOs in a 3D space, configuration of such closed systems is quite complex and cannot be explained or predicted. It also depends on the metric, i.e. on how the similarities (affinities, proximities, differences, etc.) between a drifter and fixed objects are established, which, however, is not a principal issue at this stage of investigation of this phenomenon and is interesting only from the standpoint of establishing the general regularities upon the unlimited expansion of the number of possible variants of HSMOR`s.

Regardless of how the contemporary science interprets the notion of 'attractor' [14], the aggregates of points of transition between different kinds of grouping can be referred to as attractors, or, more precisely, closed spatial attractor membranes. Their position may be coaxial, their shape can dramatically vary upon the slightest variations in the values of parameters describing the fixed objects, they may be adjacent or overlapping, but, as a rule, the intra-membrane spaces separated by them never merge.

As the attractor membranes separate different kinds of hierarchical grouping between the objects, and each of the opposite surfaces of a membrane is facing a different kind of grouping (which may be a very complex multi-node structure, especially in systems with a large number of FOs), their structure should certainly be complex. As was found in these experiments, the internal structure of attractor membranes is indeed quite complex. For instance, the thickness of membrane's subsurface areas may greatly vary. We have developed a special mathematical tool for investigation of the fine internal structure of attractor membranes, which provides extremely accurate quantitative measurements. Due to non-triviality of the said technique and the complexity of interpretation of the results, its description and discussion deserve a separate publication in the future. However, a certain level of investigation into the properties of attractor membranes has been possible by using the iterative averaging method. For instance, it was quite easy to measure the thickness of the intra-membrane area (IMA) that separates two subsurface areas of the membrane which are responsible for two different kinds of grouping. The IA-method allows absolutely reproducible measurements of IMA with an accuracy of up to the $20^{th}$ decimal place. The IMA thickness appeared to be vanishingly small, and a change of its value by 1 at any decimal place resulted in a transition to another kind of grouping. In another experiment (FO = 2; ED metric), we measured the attractor membranes emerging in the course of the drifter's movements at a linear trajectory. The results of the experiment are demonstrated in Fig. 12, where a dotted line shows the drifter's movement. It appeared that whenever the drifter came into contact with an attractor membrane, it continued its free movement along the membrane surface, i.e. the membrane "kept" the drifter on its surface and did not let it "wander" in the inter-attractor space. The only exception was the lens-shape area (see Figs. 1 and 4b) located between the HSMOR's larger and smaller hemispheres which represented the 'AB – Dr' grouping, where the drifter did not stick to the attractor membranes and was freely crossing the HSMOR, demonstrating the "tunnel effect".

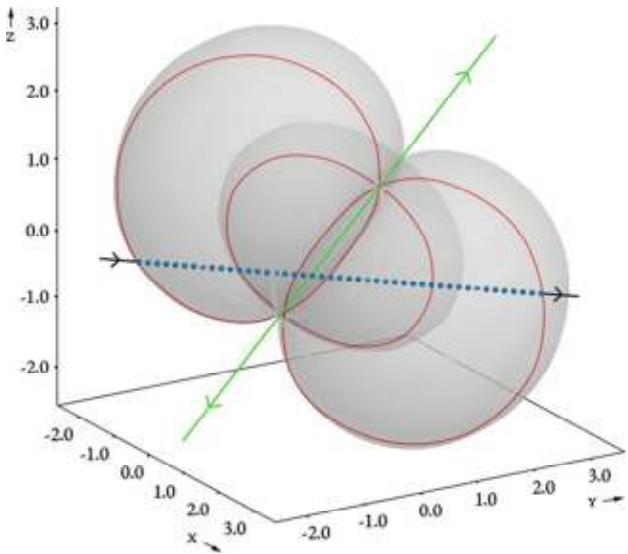

Fig.12. HSMOR(ED) represented by two fixed points A and B with the same coordinates as in Fig. 1. A dotted line shows the linear trajectory of the drifter's movement. The red lines show the trajectories of the drifter after it came into contact with attractor membrane surfaces. The green arrow shows the drifter's trajectory in the lens-shape area between the larger and smaller hemispheres of the HSMOR(ED).

## 8. Aura

In the above-described experiments, the drifter was moving chaotically and freely, i.e. neither the direction nor distance of its movement was restricted; however, in all of the experiments, it appeared that the HSMOR within which the drifter had systemic interactions with FOs represented a distinctly defined space that exceeded the actual space of the fixed objects by several times. Thus, an HSMOR could be referred to as "aura". The word aura, that came into modern usage from theosophical, occult and esoteric teachings with which this work has nothing to do, is a good term to describes the phenomenon of HSMOR: in this context, aura is that which extends beyond an actual space occupied by a system of objects and which can be discovered only through iterative averaging of the objects' properties. In all of the variants of simulation of systemic interactions between Dr and FOs described in this paper, the grouping that occurred outside of the aura was the same: a drifter was always separated, at the root-level of a hierarchical tree, from all of the fixed objects.

Our previously described method of evolutionary transformation of similarity matrices [1, 2] allows a thorough investigation of both theoretical and applied aspects of the phenomenon of aura. With all the diversity of possible configurations of fixed objects and the resulting auras, the process of formation of aura is both reproducible and regulated. For instance, a change of the coordinates of points A and B from 0 and 1 to 10 and 11, respectively, in the example illustrated in Fig. 4b, resulted in formation of aura whose size and internal structure were absolutely identical to the previous one. However, one should not overlook the problem of dialectics of similarities and differences, which inevitably arises because the iterative averaging causes two concurrent processes: consolidation of objects into two subgroups and separation into two subgroups. If formation of aura is a direct result of consolidation, the effect of separation, too, needs to be considered as these two processes occurring upon evolutionary transformation of (dis)similarity matrices are inseparable and represent a dialectical unity.

As was previously shown, we denote similarity coefficients between two subgroups emerging as a result of iterative averaging as $\Omega$ and use them for computation of angles between the branches of hierarchical trees, wherein the branch lengths are computed as natural logarithms of the numbers of iterative averaging operations required for formation of respective subgroups [1, 2]. Since a drifter's location can change from $-\infty$ to $+\infty$, the $\Omega$ values outside of an aura should tend to $-\infty$ as the drifter infinitely moves away from the fixed objects. In other words, outside of an aura, there is no connection between the aura and $\Omega$: at a certain point of the drifter's movement away from the fixed objects, the aura stops existing, whereas $\Omega$ continues to exist and has perfectly concrete values. However, within the boundaries of an aura, changes in the $\Omega$ values are extremely complex.

Dissimilarities between subgroups emerging in the course of IA can be measured as $-\ln \Omega$. The higher is the $-\ln \Omega$ value, the greater is the difference between two subgroups. It should be interesting to quantitatively analyze two areas of the changing -ln Ω values: inside and outside of an aura. Inside an aura, the changes in -ln Ω values have an extremely complex pattern even in case of a most simple system of two FOs. For understandable reasons, this coefficient equals 0 in the center of any attractor membrane at the point of transition between two kinds of grouping. Each of the HSMOR attractor membrane-enclosed areas has a specific distribution of -ln Ω values within that area. Fig. 13a shows how the -ln Ω value changes along the surface of a cross-section of HSMOR(ED) at $z = 0.5$, FO = 2 (cf. Fig. 4b). As is seen, even in case of a most simple system, visualization of the -ln Ω distribution inside of the aura is quite a complicated task.

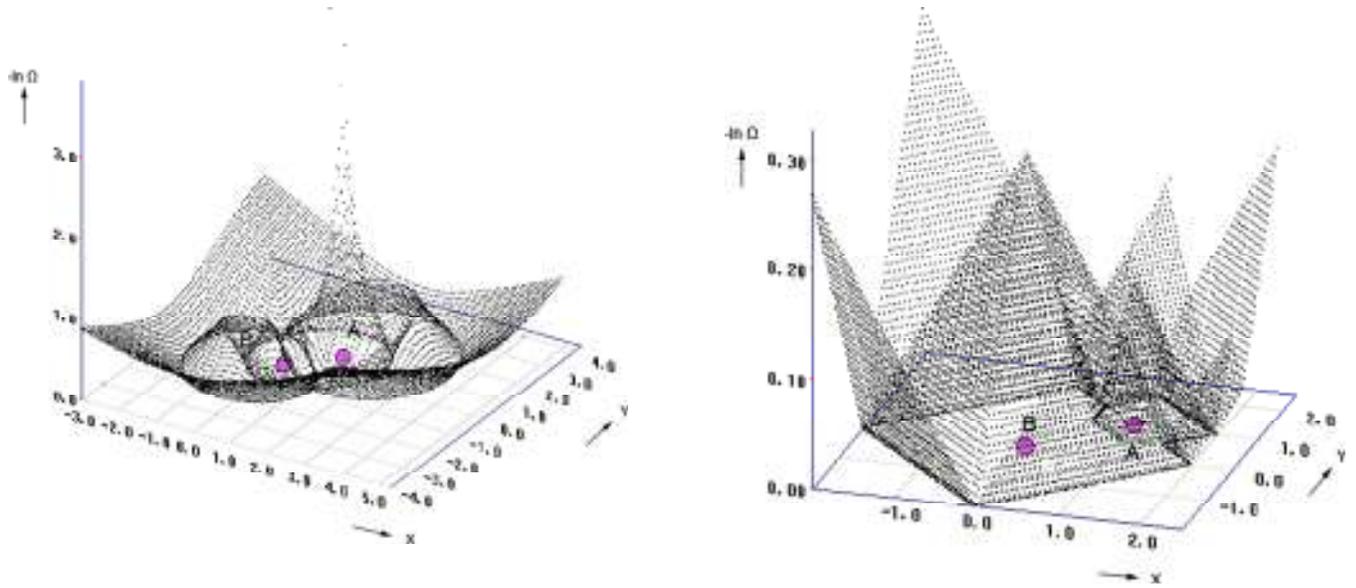

Figs.13. 13a (left): Changes in the intergroup similarity (Ω) along the surface of a cross-section of HSMOR(ED) at $z = 0.5$. Coordinates of fixed points A and B are the same as in Fig. 1. 13b (right): Same for HSMOR(XR) at $z = 2$.

The pattern of changing of the -ln Ω values largely depends on the metric. In case of the use of the XR-metric, the -ln Ω values change linearly, which is illustrated in Fig. 13b showing the changes in the -ln Ω values along the surface of the HSMOR(XR) cross-section at $z = 2$. As was shown in Fig. 6a, the cross-section at $z = 2$ represents a figure consisting of a tetragon partially overlapping with a smaller size hexagon. As is seen in Fig. 13b, the -ln Ω reaches maximum values at the center points of the tetragon and hexagon and linearly drops to 0 in the attractor

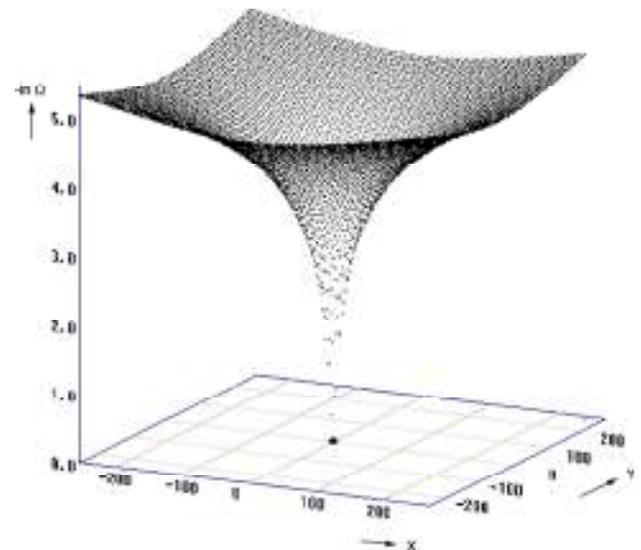

Fig. 14. Changes in the intergroup similarity (Ω) upon the drifter's moving away from fixed objects A and B. Coordinates of fixed points A and B are the same as in Fig. 1. The ED metric.

membrane areas that are hedging the two polyhedrons. In complex auras, the changes in -ln Ω values upon the use of the XR-metric are manifested in the form of a multitude of overlapping pyramids of different heights. As to the -ln Ω values outside of the aura, it seems that they change indefinitely so far as the drifter moves away from the FOs. This is illustrated in

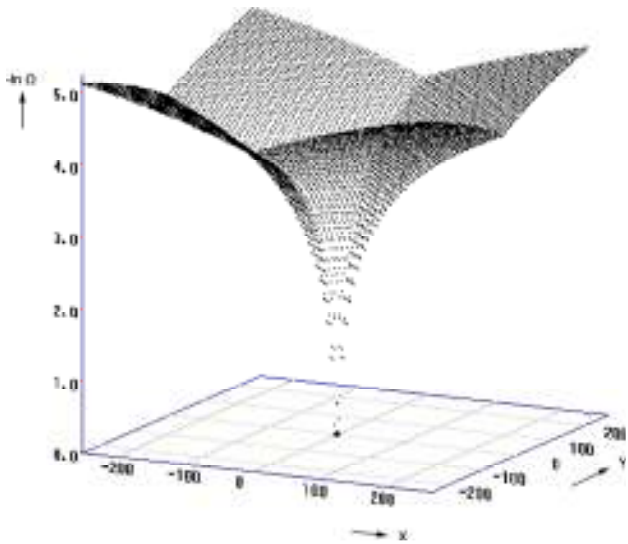

Fig. 15. Changes in the intergroup similarity ($\Omega$) upon the drifter's moving away from fixed objects A and B. Coordinates of fixed points A and B are the same as in Fig. 1. The CB metric.

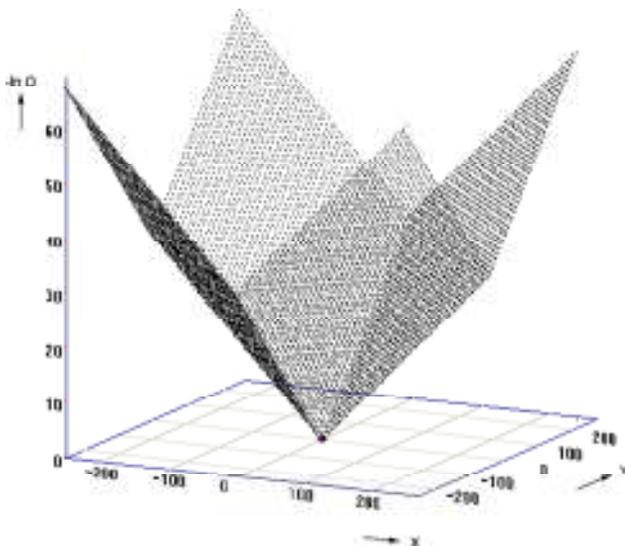

Fig. 16. Changes in the intergroup similarity ($\Omega$) upon the drifter's moving away from fixed objects A and B. Coordinates of fixed points A and B are the same as in Fig. 1. The XR metric.

figs. 14, 15 and 16 which show the changes in the -ln $\Omega$ values within an area that exceeds the area occupied by FOs by a quarter-million times. Upon the use of the XR-metric, changes in the -ln $\Omega$ values occur linearly (Fig. 16), unlike the case with two other metrics. Thus, a smooth and logically predetermined changing of the -ln $\Omega$ values upon the drifter's infinite distancing itself from the fixed objects once more emphasizes the reality and uniqueness of the emergence of a discrete aura of a transcendental nature, which occurs upon simulation of systemic processes described in this paper.

## 9. Conclusions

The results reported in this paper can serve as a convincing illustration proving the preponderance of the paradigm of holism over any other approaches based on isolation and investigation of individual parts of a whole. They also show that the superfluous usage of terms 'system', 'systemic approach', etc. in contemporary science, as well as the attempts of deeper logical understanding of the phenomena referred to by these terms have very little, if anything at all, to do with the actual phenomenon of systemic cooperative interactions that universally exist in both living and non-living nature. Cooperative interactions in systems cannot be revealed or understood by using the approaches traditionally employed by science for that purpose. As has been shown in this work, simulation of a system formation from even a minimal number of elements provides results whose scientific interpretation is *a priori* difficult if not impossible. Essentially, a holistic space of multi-object relations, visualized with the help of the algorithm of iterative averaging, turns into a tangible reality by means of quite simple mathematical operations for computer simulation on an ordinary PC. The only thing that looks extraordinary here is the concept of the isotropic field of one of the system's elements that comes into interaction with the rest of the system's elements. As we pointed out earlier, such an isotropic field could serve as one of possible implementations of the concept of computer ego. However, based on the experimental material provided in this paper, there is no direct indication that isotropic fields can vary qualitatively and quantitatively, which is certainly in conflict with the concept of ego. Therefore, the following remarks are necessary.

The methods and techniques provided in this work along with the examples illustrating the feasibility of its practical application is only a basic methodology. The next step should be the development of technology for modification of a drifter's behavior and selective sensitivity. This can be achieved through the use of coefficients that will enable a drifter to assess the importance of certain parameters. This could be used in medical diagnostics and statistics by making a drifter "focused" on searching for specific targets. Clearly, the pre-programming of a drifter's response will involve very complex mathematical operations. Another interesting applied aspect of this methodology is investigations of HSMOR in conditions of spatial-temporal variations in locations of fixed objects, which, too, will require the development of a by far more complex computer technology.

It is also important to note that HSMOR can be analyzed in spaces described by tens or hundreds of variables and that nevertheless the main results can be viewed in 2D and 3D spaces. In future publications, we will disclose some new technologies that can make HSMOR investigations more affordable and efficient.

**Acknowledgments**
Software implementation of the presented methodology has been done in collaboration with Oleg Rogachev. Graphic visualization work by Michael Andreev.

**References**
[1] L. Andreev. Unsupervised automated hierarchical data clustering based on simulation of a similarity matrix evolution. U.S. Patent 6,640,227 (2003).
[2] L. Andreev. From a set of parts to an indivisible whole. Part I. Operations in a closed mode. arXiv:0803.0034 [cs.OH] (2008).
[3] V. Estivill-Castro. *SIGKDD Explorations*, Vol.4, Issue 1, p.65 (2002).
[4] L. Andreev. From a set of parts to an indivisible whole. Part II. Operations in an open comparative mode. arXiv:0805.0455 [cs.OH] (2008).
[5] L. Andreev and M. Andreev. Method and computer-based system for non-probabilistic hypothesis generation and verification. U.S. Patent 7,062,508 (2006).
[6]. G. J. Klir. An approach to general systems theory. New York, van Nostrand (1970).
[7] M. Bunge. Treatise on basic philosophy, Vol.4, Ontology II: A world of systems. Dordrecht, Reidel (1979).
[8]. V. Majernik. Systems theoretical approach to the concept of organization. In: Altmann, G., Koch, W.A. (Eds), Systems. New Paradigms for the Human Sciences. p.126, Walter de Gruyter, Berlin, New York (1998).
[9]. S. Watanabe. Knowing and guessing. Wiley & Sons, New York (1969).
[10]. I. Prigogine, G. Nicolis. Self-Organization in Non-Equilibrium Systems. Wiley & Sons, New York (1977).
[11]. H. Haken. Synergetik. Springer-Verlag, Berlin, Heidelberg, New York (1982).
[12] E. F. Krause. Taxicab Geometry. Dover Publications (1987).
[13] L. Andreev. High-dimensional data clustering with the use of hybrid similarity matrices. U.S. Patent 7,003,509 (2006).
[14] J. Milnor. On the concept of attractor. *Communications of Mathematical Physics* 99:177-195 (1985).